\documentclass[aip,rsi,amsmath,amssymb,reprint,]{revtex4-1}

\pdfoutput=1

\usepackage{graphicx}
\usepackage{dcolumn}
\usepackage{bm}

\usepackage[utf8]{inputenc}
\usepackage[T1]{fontenc}
\usepackage{mathptmx}
\usepackage[separate-uncertainty=true]{siunitx}
\usepackage{tabularx}
\usepackage[dvipsnames]{xcolor}

\newcommand{\twoDEG}{2D electron gas}
\newcommand{\QD}{quantum dot}
\newcommand{\PTR}{pulse tube refrigerator}

\begin{document}

\title{Quantum dot thermometry at ultra-low temperature in a dilution refrigerator with a $^4$He immersion cell}

\author{G. Nicol\'i}
\affiliation{Solid State Physics Laboratory, ETH Z\"urich, Otto-Stern-Weg 1, 8093 Z\"urich, Switzerland}

\author{P. M\"arki}
\affiliation{Solid State Physics Laboratory, ETH Z\"urich, Otto-Stern-Weg 1, 8093 Z\"urich, Switzerland}

\author{B. A. Br\"am}
\affiliation{Solid State Physics Laboratory, ETH Z\"urich, Otto-Stern-Weg 1, 8093 Z\"urich, Switzerland}

\author{M. P. R\"o\"osli}
\affiliation{Solid State Physics Laboratory, ETH Z\"urich, Otto-Stern-Weg 1, 8093 Z\"urich, Switzerland}

\author{S. Hennel}
\affiliation{Solid State Physics Laboratory, ETH Z\"urich, Otto-Stern-Weg 1, 8093 Z\"urich, Switzerland}

\author{A. Hofmann}
\affiliation{Solid State Physics Laboratory, ETH Z\"urich, Otto-Stern-Weg 1, 8093 Z\"urich, Switzerland}

\author{C. Reichl}
\affiliation{Solid State Physics Laboratory, ETH Z\"urich, Otto-Stern-Weg 1, 8093 Z\"urich, Switzerland}

\author{W. Wegscheider}
\affiliation{Solid State Physics Laboratory, ETH Z\"urich, Otto-Stern-Weg 1, 8093 Z\"urich, Switzerland}

\author{T. Ihn}
\affiliation{Solid State Physics Laboratory, ETH Z\"urich, Otto-Stern-Weg 1, 8093 Z\"urich, Switzerland}

\author{K. Ensslin}
\affiliation{Solid State Physics Laboratory, ETH Z\"urich, Otto-Stern-Weg 1, 8093 Z\"urich, Switzerland}

\date{\today}

\begin{abstract}
Experiments performed at a temperature of a few millikelvin require effective thermalization schemes, low-pass filtering of the measurement lines and low-noise electronics. Here, we report on the modifications to a commercial dilution refrigerator with a base temperature of \SI{3.5}{\milli\kelvin} that enable us to lower the electron temperature to \SI{6.7}{\milli\kelvin} measured from the Coulomb peak width of a quantum dot gate-defined in an [Al]GaAs heteostructure. We present the design and implementation of a liquid $\mathrm{^4He}$ immersion cell tight against superleaks, implement an innovative wiring technology and develop optimized transport measurement procedures.
\end{abstract}

\maketitle

\section{\label{sec:intro}Introduction}

Quantum phenomena in solid state nanostructured systems are nowadays investigated using commercially available dilution refrigerators that enable experiments at temperatures below \SI{10}{\milli\kelvin}. Energy resolution on the few \si{\micro\electronvolt} scale led to the discovery of novel states of matter as fractional quantum Hall states,\cite{stormer_fractional_1983,nayak_non-abelian_2008-1} non-trivial superconductivity\cite{cao_unconventional_2018,mackenzie_superconductivity_2003} and topological insulators.\cite{hasan_colloquium_2010} However, a low mixing chamber temperature does not necessarily result in an equivalently low electron temperature $T_\mathrm{e}$, due to vanishingly low electron-phonon coupling and residual heat leaks, which become relevant at low temperatures.\cite{wellstood_hot-electron_1994,giazotto_opportunities_2006,roukes_hot_1985}

Efficient filtering and thermal anchoring of the measurement lines are crucial steps towards minimizing $T_\mathrm{e}$, but these alone do not ensure optimal thermalization of the measured devices. Significant improvements have been obtained using liquid $\mathrm{^3He}$ immersion cells to cool down both the device and the measurement leads.\cite{xia_ultra-low-temperature_2000,huang_disappearance_2007,samkharadze_integrated_2011,knighton_evidence_2018,pan_exact_1999} This success comes with the increased complexity of handling $\mathrm{^3He}$, an element that is rare and expensive.\cite{cho_helium-3_2009} Adiabatic nuclear demagnetization\cite{lounasmaa_experimental_1974,clark_method_2010} also proved to be a powerful tool in reaching record low temperatures,\cite{tuoriniemi_nuclear_2000,palma_magnetic_2017,palma_-and-off_2017,yurttagul_indium_2018} with recent results demonstrating electron temperatures in a network of metallic single electron transistors in the hundreds of \si{\micro\kelvin} regime and hold times in the range of tens of hours.\cite{sarsby_500_2019-1}

While reaching low electron temperature is a challenge in itself, directly measuring a $T_\mathrm{e}<\mathrm{\SI{10}{\milli\kelvin}}$ requires specifically designed systems. The temperature dependence of nanoelectronic devices like metallic single-electron transistors has been widely used.\cite{iftikhar_primary_2016} Other approaches involve using the temperature dependence of shot noise through quantum point contacts\cite{iftikhar_primary_2016,bid_shot_2009,venkatachalam_local_2012} or of the current through metal-superconductor tunnel junctions.\cite{feshchenko_tunnel-junction_2015} Coulomb blockade thermometers, a network of metallic islands separated by tunnel junctions,\cite{knuuttila_direct_1998,casparis_metallic_2012} have emerged as the premium choice for direct measurement of ultra-low electron temperatures.\cite{bradley_nanoelectronic_2016,sarsby_500_2019-1} Such devices can be designed to have improved electron-phonon coupling and resilience to noise and sample specific details. The disadvantage is that ultra-low electron temperatures in these metallic thermometers do not necessarily correspond to the same temperatures in samples made of different materials, such as semiconductors, and with different structures.

In this manuscript, we present several custom-made modifications performed on a commercial dilution refrigerator. Among these, we designed and built a superfluid $\mathrm{^4He}$ immersion cell, including suitable filling and sealing systems. We discuss design choices and operation procedures that lead us to the optimization of our setup.

Subsequently, we describe our electron temperature measurements based on a home-built quantum dot primary thermometer electrostatically defined within an [Al]GaAs heterostructure.\cite{kouwenhoven_few-electron_2001,ciorga_addition_2000} This is the same material used to perform other experiments such as the investigation of fractional quantum Hall states\cite{stormer_fractional_1983,nayak_non-abelian_2008-1} or the interference of quasi-particles.\cite{nakamura_aharonovbohm_2019} We argue that the estimate for $T_\mathrm{e}$ can be realistically extended to other devices. We describe several optimization steps performed on the measurement setup that led us to measure a lowest $T_\mathrm{e}=\mathrm{\SI{6.74\pm{}0.07}{\milli\kelvin}}$ via suitable measurement procedures.\cite{iftikhar_primary_2016,ihn_semiconductor_2010,maradan_gaas_2014}

\section{\label{sec:fridge}Cryogenic setup and $^4$He immersion cell}

\begin{figure*}
\includegraphics{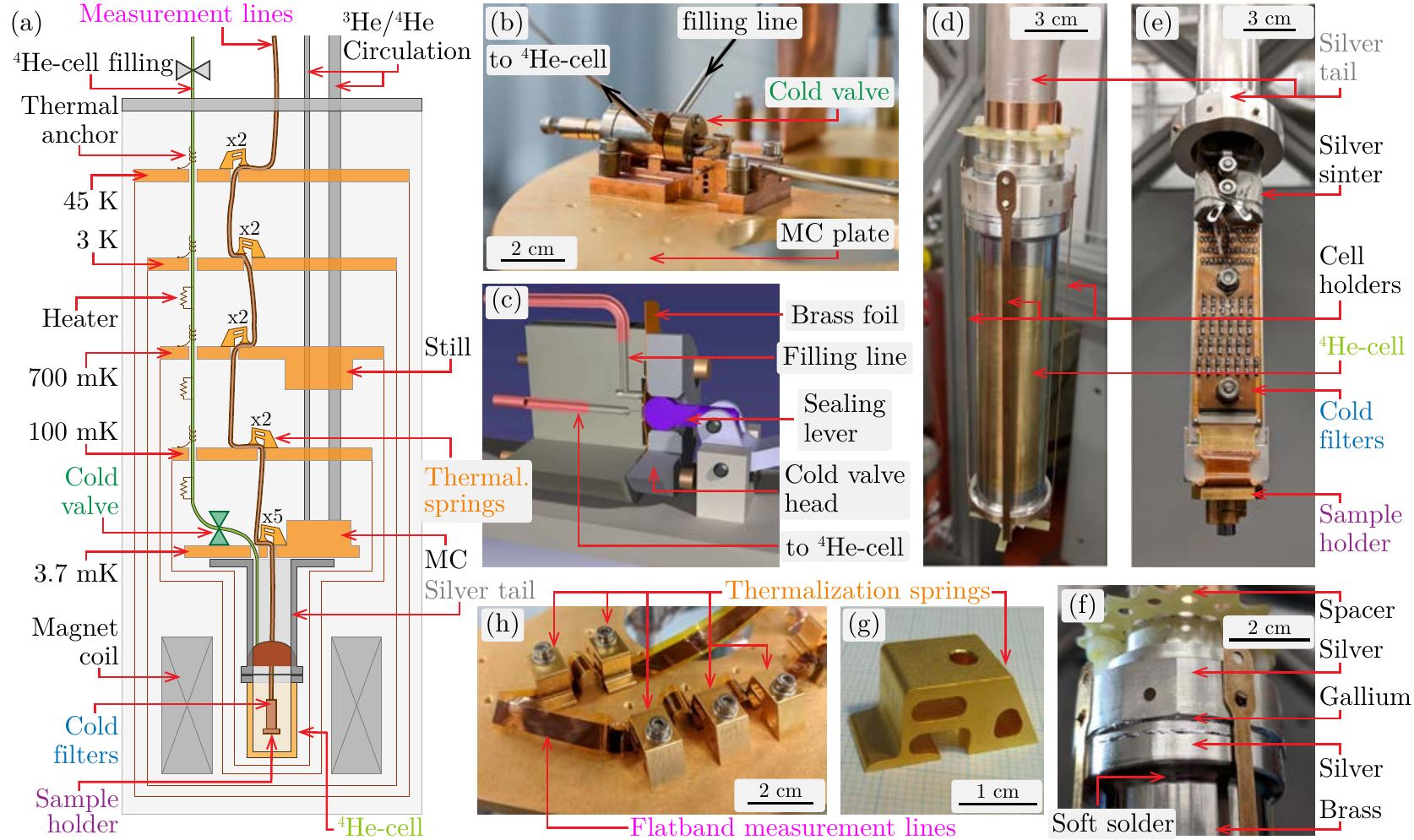}
\caption{\label{fig1}(a) Schematics of the dilution refrigerator modified with a $\mathrm{^4He}$ immersion cell (not to scale). The main components and the characteristic temperatures of the setup are indicated. (b) Photograph of the cold valve sitting on the mixing chamber plate that separates the volume of the immersion cell from the filling line. (c) Schematic drawing of the head of the cold valve showing the sealing mechanism. A metallic cylinder (purple) presses a brass foil (orange) against the body of the valve, separating the input and output capillary lines (red). (d,e) Photographs of the bottom of the silver tail both with (d) and without (e) the immersion cell. The chip-carrier is pressed against the tail by a metallic spring and is connected to filters that are cooled to the same temperature. (f) Photograph of the contact point between the immersion cell and the silver tail. The sealing is ensured by a gallium soldering. The material of the parts is indicated. (g) Metallic spring used to thermally anchor the measurement wires to the cryostat. (h) Zoom-in on part of the mixing chamber plate where the custom-made flatband measurement wires are fixed to the plate by the springs shown in panel (g).}
\end{figure*}

The setup presented in this manuscript is based on an Oxford Instruments cryogen-free dilution refrigerator. Two pulse tube compressors provide the cooling power necessary to reach a low enough temperature for the dilution unit to operate. Thus, no refill of liquid coolants is required for the continuous operation of the setup, at the cost of mechanical vibrations generated by the switching valve of the compression system. A sketch of the setup is shown in Fig.~\ref{fig1}(a). The base temperature of the mixing chamber reaches values below \SI{4}{\milli\kelvin} at zero magnetic field, as reproducibly measured with a commercial noise thermometer.\cite{engert_practical_2009} We have observed an increase in base  temperature of \mbox{\SIrange[range-phrase = -- , range-units = single]{2}{4}{\milli\kelvin}} when a field of $\sim\mathrm{\SI{10}{T}}$ is applied. The experiments discussed here are all performed without magnetic field.

In order to improve the quality of our measurements (lower electronic temperature of the devices, better signal-to-noise ratio), we introduced several modifications to the commercial setup: custom-made wires with reworked thermal anchoring to the different stages of the refrigerator and cold filters built in-house are used and low-noise, custom-made voltage sources and IV-converters are employed to perform electronic transport measurements. Furthermore, we engineered a superfluid $\mathrm{^4He}$ immersion cell to improve the thermalization of the measured devices and the mixing chamber plate. This endeavor required the development of a pipe and valve system to fill and successively empty the immersion cell and to seal it during normal operation. Next, we will discuss all of these custom-made modifications, starting from the latter.

\subsection{Immersion cell and cold valve design}

At approximately \SI{2.2}{\kelvin} and normal pressure, liquid $\mathrm{^4He}$ undergoes a phase transition and becomes superfluid.\cite{wilks_introduction_1987,pobell_matter_1992} In this state the liquid flows without viscosity and is unable to sustain a temperature gradient, thus having a uniform temperature. Thermal conductivity becomes virtually infinite, under the conditions that temperature or fluid velocity do not exceed some critical values.\cite{wilks_introduction_1987,pobell_matter_1992} These properties make superfluid $\mathrm{^4He}$ an appealing candidate to improve the thermal coupling of mesoscopic devices and measurement lines to the refrigerator's cooling power.

On the other hand, dealing with a superfluid poses several challenges concerning its confinement in closed volumes. A film of liquid $\mathrm{^4He}$ will cover all the surfaces available and even the tiniest fracture in the confining material will allow it to escape. Therefore, the first step was to develop a valve suitable for operation with superfluid $\mathrm{^4He}$ and capable of switching its status at base temperature while remaining superleak-tight. As will be explained in section~\ref{ssec:filling}, such a component is crucial for the operation of the immersion cell setup. A picture of our valve is shown in Fig.~\ref{fig1}(b), with the schematic cross section in Fig.~\ref{fig1}(c). A mechanical lever that can be pulled from outside the cryostat presses a polished brass foil against the body of the valve. The metal profile of the head of the valve separates the inlet and outlet $\mathrm{^4He}$ lines, as shown in Fig.~\ref{fig1}(c). The clean metal-metal interfaces, the edged profile and the applied forces make the valve tight even against superfluid leaks while enabling mechanical switching at millikelvin temperatures.

The immersion cell itself is a hollow cylinder made of brass [Figs.~\ref{fig1}(d,e)]. The metal of the liquid $\mathrm{^4He}$ container provides shielding against electromagnetic radiation heating, at the cost of potentially introducing dissipative eddy currents while sweeping the magnetic field. To minimize this, we chose a metal alloy with high resistivity. In future iterations this could be further improved by using copper-beryllium or German silver as possible alternatives. The cell is held against the bottom end of the silver tail with copper-beryllium bands that act as springs, pressing the two parts together in vertical direction. The top part of the immersion cell is made of silver [Fig.~\ref{fig1}(f)], like the part of the tail it is in contact with. This prevents that the two parts slide against each other due to thermal shrinking during cool-down, which could potentially create leaks at base temperature.

Sealing the immersion cell against the silver tail is not only achieved by pressing the two parts together with the spring holders, but also by soldering them with gallium. A layer of high-purity Ga is spread on the two silver surfaces while warming them up to $\sim\mathrm{\SI{40}{\celsius}}$ to improve their wettability. We then connect the two parts and let the liquid metal cool down and solidify. Figure~\ref{fig1}(f) shows a picture of the sealed cell. Gallium is known to diffuse into silver and form an alloy with it.\cite{baren_ag-ga_1990} We believe that this phenomenon helps to create a superleak-tight seal for the immersion cell. The observation of the Ag-Ga alloy formation is further supported by the fact that in order to disjoin the soldering, it is not enough to heat the cell above the melting point of gallium ($\sim\mathrm{\SI{29.7}{\celsius}}$) for an extended period of time. Depending on the relative concentration of the two elements, the melting temperature of the alloy is much higher.\cite{baren_ag-ga_1990} Therefore, we used shear force between the cell and the tail applied with a tool appositely built to get them apart, while again heating to $\sim\mathrm{\SI{40}{\celsius}}$.

We confirmed the tightness of the whole filling line-cold valve-immersion cell system by performing leak tests with $\mathrm{^4He}$ both at room temperature and at base temperature. Cold leaks between the immersion cell system and the cryostat vacuum chamber can be detected as a sudden increase in temperature of the different stages of the refrigerator due to the presence of exchange gas in the vacuum chamber. With the absence of this observation in our final setup, the whole system supports superfluid $\mathrm{^4He}$ operation for an extended period of time. The experiments described in this manuscript run for about 5 weeks, after which we emptied the cell and warmed up the setup without experiencing any leak.

Inside the immersion cell, the sample sits at the very bottom of the silver tail, as shown in Fig.~\ref{fig1}(e). A metallic sample holder covers the chip-carrier and is screwed to the tail. The holder shields the sample from external electromagnetic radiation. Its spring-like design presses the chip-carrier against the the tail, ensuring the effective cooling of the former. The carrier is electrically connected to the cold filters, which are also cooled down in the immersion cell, minimizing the Johnson noise produced by the resistances of the filters and participating in the cooling of the measurement lines.

Kapitza resistance is a known hindrance to the thermalization between a liquid and a solid.\cite{pobell_matter_1992,wilks_introduction_1987} In the $\mathrm{^4He}$ cell silver sinter foils are stacked together and then screwed to the silver tail to ensure their thermalization. The high surface-to-volume ratio of this material enables a large contact area ($\sim\mathrm{\SI{30}{\metre^2}}$) with the superfluid $\mathrm{^4He}$ present in the cell even with little space available. Increasing the contact area between the bath and the cold metal via the silver sinter proved to be effective in cooling our samples as witnessed by the electronic temperature measurements shown in section~\ref{sec:QD}.

\subsection{\label{ssec:filling}$^4$He cell filling operation}

Filling $\mathrm{^4He}$ into the immersion cell starts from a bottle of gas kept at room temperature. We use two buffer volumes to control the pressure at which the gas is let into the cold part of the system and then into the cell, where it condenses. A too high pressure would introduce too much heat in the cryostat, leading to an increase in still and mixing chamber temperatures beyond the point at which the dilution unit can operate. On the other hand, a too low pressure increases the time required to fill the immersion cell with superfluid $\mathrm{^4He}$. As a compromise, we keep a pressure of \mbox{\SIrange[range-phrase = -- , range-units = single]{50}{60}{\milli\bar}} during the condensation operation: first a smaller volume ($\mathrm{\SI{4.5}{\litre}}$) is filled directly from a pressurized bottle of $\mathrm{^4He}$ up to a pressure of $\sim\mathrm{\SI{1}{\bar}}$, then the second, larger volume ($\mathrm{\SI{45}{\litre}}$) is filled with gas up to the condensing pressure.

The second buffer volume is connected to the inner part of the cryostat and to the immersion cell through a thin stainless steel capillary. We used pipes of different sizes soldered together, whose inner diameter ranged from \SI{1}{\milli\meter} to \SI{0.8}{\milli\meter}. The capillary is thermally anchored to all the plates of the cryostat in order to gradually reduce its temperature until it reaches the mixing chamber plate. Thermal coupling is provided by flexible braided copper wires, which are at one end wound and soldered to the filling line and at the other end screwed and pressed against the plates of the cryostat, as sketched in Fig.~\ref{fig1}(a). We observed the same base temperature of the cryostat before and after this addition.

After having filled the $\mathrm{^4He}$ cell, a superfluid film covers the inner surface of the filling line up to a certain height, thermally coupling higher temperature stages to the low temperature part of the setup. In this situation, it is not possible to reach the base temperature of the refrigerator and it is necessary to empty the filling line. Simply pumping on it is not enough, as the vapor pressure of liquid $\mathrm{^4He}$ quickly becomes vanishingly small at temperatures below \SI{0.5}{\kelvin}.\cite{pobell_matter_1992} It is at this point that the valve becomes crucial: by closing it we can separate the capillary used to fill $\mathrm{^4He}$ from the volume of the cell. We use several resistors sitting directly on the filling line to heat it up locally (see sketch in Fig.~\ref{fig1}(a), total dissipated power around \SI{5}{\milli\watt}) while pumping on it for around half a day. During this procedure we keep the cold valve closed, preventing the immersion cell to also be pumped on. After switching the heaters off, we let the refrigerator cool down until it reaches base temperature.

\subsection{Custom-made measurement line: design and thermalization}

Instead of using braided wires or coaxial lines for the sample cabling, we developed a home-made flatband cable.\cite{bram_imaging_2018} It consists of 49 lines made of \SI{100}{\micro\meter}-diameter manganin wires insulated by a polyimide layer. Every second of these lines is grounded at the top of the cryostat. This leads to reduced capacitive cross-talk between the 24 measurement lines which is not as perfect as with coax cables, but comes with the advantage that the flatband cable can be mechanically pressed for thermal anchoring.

Good thermal coupling between two metals requires a large force to be applied between them.\cite{ekin_experimental_2006} The design of the flatband cables allows us to do so without damaging them or breaking the insulation between measurement lines and ground lines. We use custom-built copper-beryllium springs to thermally anchor the measurement wires to the cryostat [Fig.~\ref{fig1}(g)]. Thanks to their design, these springs apply a constant force of \SI{500}{\newton} per spring over the entire section of the flatband cable they cover. We screw the springs directly to the metallic plates of the refrigerator, as shown in Fig.~\ref{fig1}(h). By doing so at all stages of the cryostat, the temperature of the flatband gradually reduces until it reaches the mixing chamber plate temperature.\cite{bram_imaging_2018,ekin_experimental_2006}

The wires of the flatband are then connected individually to the cold filters. These are first order RC low-pass filters with a bandwidth of \SI{10}{\kilo\hertz} and the only filtering present in our cryogenic setup. We do not employ microwave filters\cite{thalmann_comparison_2017} such as metal powder filters,\cite{scheller_silver-epoxy_2014,mueller_printed_2013} or cables designed to damp high frequency signals.\cite{zorin_thermocoax_1995} The combination of our custom-made room temperature electronics, flatband cables and thermalization springs enables us to reach state-of-the-art performance while keeping the wiring effort comparatively low.

\section{\label{sec:QD}Quantum dot thermometry}

The goal of the efforts described in section~\ref{sec:fridge} is to lower the electronic temperature of the devices cooled down in our setup below the $\sim\mathrm{\SI{15}{\milli\kelvin}}$ reached before. There are a few ways to directly measure the temperature of an electron gas in the millikelvin range.\cite{iftikhar_primary_2016,bradley_nanoelectronic_2016} We decided to use a \QD{} electrostatically defined in a \twoDEG{}\cite{kouwenhoven_few-electron_2001,kouwenhoven_electron_1997} as it represents the type of devices that we would later measure in our setup. Quantum dots constitute a form of primary thermometry in the temperature range we are interested in, but they are sensitive to several measurement parameters, such as the excitation applied to the sample and the bandwidth of the sampling procedure, and are influenced by sample and material specific factors, i.e. fluctuations in the background disorder potential and charge noise in the measured current.\cite{maradan_gaas_2014}

We will show below how our electronic temperature measurements depend on noise and filtering of the measurement lines, on the performance of the measurement devices, the ac excitation applied to the sample and the temporal duration of the measurements. All of these issues either affect the accuracy and precision of our estimated electronic temperature or directly heat the \twoDEG{}. The noise introduced by the instrumentation and the quality of the samples themselves, in terms of stability and charge fluctuations in its host material, limit in the end the accuracy of the measurements, such that our estimates can be considered upper limits on the real temperature of the \twoDEG{}. Finally, we will also discuss the effectiveness of our superfluid $\mathrm{^4He}$ immersion cell setup regarding its improvement of the electronic temperature of the device. Experiments are performed both with an empty and a full cell to compare the two cases.

\subsection{\label{ssec:setup}Standard measurement setup}

\begin{figure}
\includegraphics{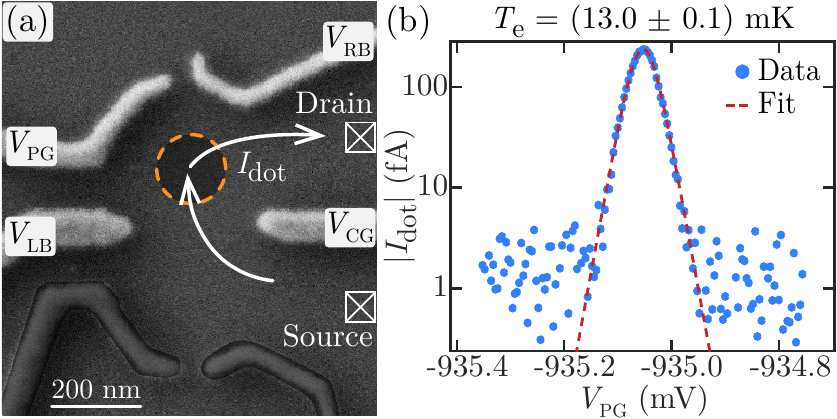}
\caption{\label{fig2}(a) Scanning electron micrograph of the sample. The light-grey features are the metallic top gates used to electrostatically define a quantum dot. Negative voltages labeled as in the figure are applied to the respective gates to locally deplete the \twoDEG{}. The dark grey gates have been kept at a constant potential and are not relevant for the discussed experiments. The dashed circle schematically represent the position where the \QD{} is formed. The white crossed boxes indicate the source and drain ohmic contacts to the \twoDEG{}. The sample is the same as the one in Refs.~\onlinecite{hofmann_measuring_2016,hofmann_anisotropy_2017}. (b) Current $I_\mathrm{dot}$ flowing through the device as a function of the  plunger gate voltage $V_\mathrm{\textsc{pg}}$ (step S1, see table~\ref{table}). The blue dots represent the data, while the red dashed line is a fit with a thermally broadened Coulomb peak function (see discussion in the text). The data is the result of an average of 50 measurements of the same peak. The dc offset current of the IV-converter is subtracted from the measurements.}
\end{figure}

The \twoDEG{} hosted by a standard modulation-doped GaAs/AlGaAs heterostructure can be locally depleted using metallic top-gates that act as Schottky contacts. A scanning electron micrograph of our sample is shown in Fig.~\ref{fig2}(a): the \QD{} is electrostatically defined using 4 different gates, allowing us to independently tune the number of electrons confined in the dot and its tunnel coupling to the source and drain leads.\cite{de_franceschi_electron_2001} The sample is the same as in Refs.~\onlinecite{hofmann_measuring_2016,hofmann_anisotropy_2017}.

We depleted the dot in the Coulomb-blockade regime down to its last electron using a quantum point contact located nearby as a charge detector to check the state of the dot (not shown in the figure).\cite{dicarlo_differential_2004} Quantum dot thermometry requires (i) conduction through the dot to happen in the single-level transport regime, (ii) tunnel coupling broadening $\Gamma \ll k_\mathrm{\textsc{B}}T_\mathrm{e}$ and (iii) source-drain bias smaller than $4k_\mathrm{\textsc{B}}T_\mathrm{e}$. These requirements are usually achieved in small dots with a low occupation number.\cite{kouwenhoven_electron_1997,maradan_gaas_2014,ihn_semiconductor_2010}

We measured the current $I_\mathrm{dot}$ through the \QD{} in response to a small dc bias of \SI{1}{\micro\volt} as a function of the plunger-gate voltage $V_\mathrm{\textsc{pg}}$. The result, shown in Fig.~\ref{fig2}(b), follows the expected conductance resonance line shape\cite{kouwenhoven_few-electron_2001} and the current is strongly suppressed off-resonance. The resonance corresponds to the alignment of the dot's electrochemical potential to the bias window opened between the electrochemical potential of the source and drain electron reservoirs.

Assuming that conditions (i)-(iii) are fulfilled in the experiment, we can describe this peak with the equation\cite{ihn_semiconductor_2010}
\begin{equation}
I_\mathrm{dot} = \frac{I_\mathrm{0}}{\cosh^2 \left[ \alpha e\left( V_\mathrm{\textsc{pg}}-V_\mathrm{0}\right) /2k_\mathrm{\textsc{b}}T_\mathrm{e}\right]},
\label{eq:peak}
\end{equation}
where $e$ is the electron charge, $k_\mathrm{\textsc{b}}$ is the Boltzmann constant, $I_\mathrm{0}$ is the peak amplitude, $V_\mathrm{0}$ is the peak position in plunger gate voltage, $T_\mathrm{e}$ is the electronic temperature of the sample and $\alpha$ is the lever-arm of the plunger gate on the energy states of the dot. The lever-arm can be obtained performing finite-bias measurements known as Coulomb diamonds\cite{ihn_semiconductor_2010} and depends mostly on the gate geometry and weakly on the specific tuning of the voltages that define the dot. For all the experiments discussed here, $\alpha\sim{}0.077$, known to a statistical uncertainty of less than \SI{1}{\percent}.

The dependence of Eq.~(\ref{eq:peak}) on $T_\mathrm{e}$ allows us to directly calculate the electronic temperature of the \twoDEG{} in the device from the measured full width at half maximum of the Coulomb peak. A fit to the data with this function is also plotted in Fig.~\ref{fig2}(b), showing excellent agreement with the data and yielding $T_\mathrm{e}=\mathrm{\SI{13.0\pm{}0.1}{\milli\kelvin}}$. Considering that the base temperature of our cryogenic setup is $<\mathrm{\SI{4}{\milli\kelvin}}$, the electron gas in the sample does not fully thermalize or is additionally heated, possibly due to the vanishing electron-phonon coupling in the device, a higher temperature at the bottom of the silver tail compared to the mixing chamber plate, noise being transmitted via the measurement lines into the \twoDEG{} or a combination of these effects.

This experiment has been performed leaving the immersion cell empty and using a standard dc setup in our laboratories: custom-made digital-to-analog low-noise voltage sources have been used to energize the gates and to provide the dc bias (in this case with a $1/1000$ divider). They are also filtered at room temperature by low-pass first order RC filters with a time constant $\tau{}=\mathrm{\SI{4}{\milli\second}}$. Their output voltage noise is $<\mathrm{\SI{0.4}{\micro\volt}}_\mathrm{rms}$ in the frequency range from \SI{0.1}{\hertz} to \SI{100}{\kilo\hertz}. We measured the current using IV-converters built in-house incorporating OPA140\cite{noauthor_opa140_nodate} with a feedback resistance $R_\mathrm{\textsc{f}}=\mathrm{\SI{1}{\giga\ohm}}$.

\begin{table}
\caption{\label{table} Description of the different steps for the optimization of the electronic temperature $T_\mathrm{e}$ and quantum dot thermometry techniques. The improvements and changes performed in each step are carried over for the successive measurements. The step numbers (S1 to S6) are referred to throughout the discussion in this manuscript. The values reported in the table correspond to the averaged $T_\mathrm{e}$ at base temperature in each step. The errors presented correspond to one standard deviation $\sigma$ obtained from the statistical analysis of repeated measurements.}
{\renewcommand{\arraystretch}{1.4}
\begin{ruledtabular}
\begin{tabularx}{\columnwidth}{r  D{,}{\pm}{-1}  p{5.8cm}}
Step & \multicolumn{1}{c}{$T_\mathrm{e}$ (\si{mK})} & Description\\
\hline
\textbf{S1} & 13.0,0.1 & Standard dc measurement setup with custom-made IV-converter (\SI{8}{\nano\volt\per\sqrt{Hz}} input voltage noise @ \SI{33}{\hertz})\\
\hline
\textbf{S2} & 11.2,0.1 & Additional filtering on all gate voltages (time constant $\tau$ increased from \SI{4}{\milli\second} to \SI{22}{\milli\second})\\
\hline
\textbf{S3} & 9.49,0.03 & Ac measurement setup with two-stage low-noise IV-converter (\SI{1}{\nano\volt\per\sqrt{Hz}} input voltage noise @ \SI{33}{\hertz})\\
\hline
\textbf{S4} & 8.20,0.04 & Immersion cell filled with superfluid $\mathrm{^4He}$\\
\hline
\textbf{S5} & 7.45,0.08 & Measurement time and ac excitation amplitude $V_\mathrm{\textsc{ac}}$ optimization\\
\hline
\textbf{S6} & 6.74,0.07 & Pulse-tube refrigerators turned off\\
\end{tabularx}
\end{ruledtabular}
}
\end{table}

The measurement performed with this setup corresponds to step S1 of our optimization process, as reported in Table~\ref{table}. This table briefly describes the changes performed between each measurement and shows the electronic temperatures measured in each case. For the rest of this manuscript we will refer to the numbering of steps indicated in the table and will describe them in more detail. All of these steps were performed in the same thermal cycle of our refrigerator in order to reduce variability related both to the setup and to the device.

In step S2, we improved the filtering at the output of the voltage sources, increasing the time constant $\tau$ from \SI{4}{\milli\second} to \SI{22}{\milli\second}. Noise on the voltage applied to the gates is transmitted to the dot, which experiences a fluctuating potential, leading to fluctuations in the peak position $V_\mathrm{0}$. When sampling the peak with current measurements, the finite bandwidth of the measurement effectively averages the shifting Coulomb resonance, causing an inflated peak width and consequently an artificially inflated extracted temperature. We performed the same measurements and analysis as described in step S1 and observed that the increased filtering reduced the estimated temperature $T_\mathrm{e}$ to \SI{11.2\pm{}0.1}{\milli\kelvin}.

\subsection{\label{ssec:lownoise}Temperature reduction with low-noise instrumentation}

Noise characteristics of measurement electronics is relevant not only for the signal-to-noise ratio of measured data, but also directly acts back onto the investigated sample: the input voltage noise of amplifiers is directly transmitted to the electron gas. Independent of the measurement frequency, the sample picks up noise in the whole frequency range transmitted by measurement lines and cold filters. The cut-off frequency in our case is at \SI{10}{\kilo\hertz}. In some cases, this additional noise can increase the temperature of the electrons in a significant way.

\begin{figure}
\includegraphics{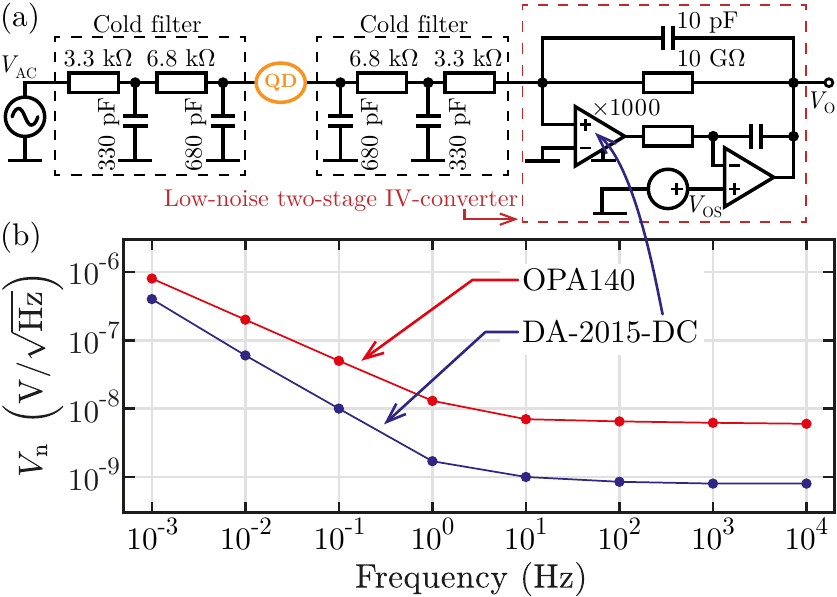}
\caption{\label{fig3}(a) Circuit schematics of the measurement setup used for steps S3--S6 (see table~\ref{table}). The red box highlights the low-noise two-stage IV-converter, while the black boxes indicate the cold filters circuitry thermalized with the mixing chamber plate and with the silver tail inside the refrigerator. The yellow circle represents the \QD{}. An ac excitation voltage is applied to the source contact. The measurement instrumentation is connected to the drain contact. An unwanted dc bias across the dot is manually compensated with $V_\mathrm{\textsc{os}}$. (b) Comparison  of the input voltage noise $V_\mathrm{n}$ of the OPA140 incorporated in the IV-converters employed in steps S1 and S2 and of the low-noise amplifier DA-2015-DC\cite{marki_temperature-stabilized_2017-1} constituting the first stage of the two-stage IV-converter used in steps S3-S6.}
\end{figure}

In step S3, we checked the influence of the IV-converter input voltage noise in our setup by performing the same measurement as in step S1 and S2 but with a new low-noise two-stage IV-converter. This device incorporates as an input stage the low-noise temperature-stabilized differential amplifier DA-2015-DC, developed in our laboratories.\cite{marki_temperature-stabilized_2017-1} Figure~\ref{fig3}(a) shows a circuit schematic of the new setup. Compared to the OPA140, the DA-2015-DC exhibits lower input voltage noise density $V_\mathrm{n}$ at all the tested frequencies, as shown in Fig.~\ref{fig3}(b).

We measure the current $I_\mathrm{dot}$ in response to an ac excitation with amplitude $V_\mathrm{\textsc{ac}}=\mathrm{\SI{500}{\nano\volt}}$ at a frequency of \SI{33}{\hertz}. The output $V_\mathrm{\textsc{o}}$ of the two-stage IV-converter is then measured with a lock-in amplifier. The time constant of the lock-in is set to \SI{100}{\milli\second} and its filter steepness to \SI{6}{dB/oct}.  At the measurement frequency, the feedback loop impedance and the gain of the IV-converter is dominated by the \SI{10}{\pico\farad} capacitor. Circuit simulations show that both the gain and the phase of the output signal are stable in the resistance range spanned by the \QD{}.

\begin{figure}
\includegraphics{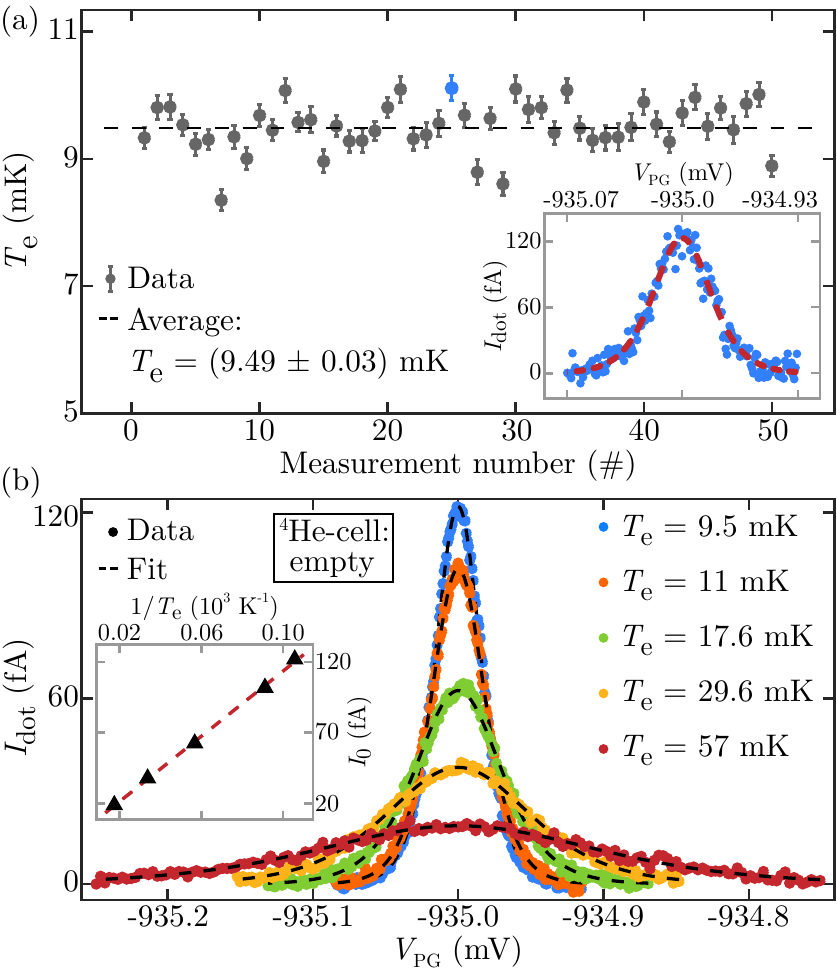}
\caption{\label{fig4}(a) Electronic temperature $T_\mathrm{e}$ obtained from repeated measurements of the Coulomb resonance in the dot current $I_\mathrm{dot}$ without additional heating. Each measurement yields a value for $T_\mathrm{e}$ and a statistical average of these values results in $T_\mathrm{e} =\mathrm{\SI{9.49\pm{}0.03}{\milli\kelvin}}$. The inset shows a representative measurement, where the fitted curve is drawn with a red dashed line and the measured data is represented with blue dots. The value for $T_\mathrm{e}$ extracted from this specific measurement is blue-colored in the main plot. (b) Dot current $I_\mathrm{dot}$ as a function of the plunger gate voltage $V_\mathrm{\textsc{pg}}$ measured at different mixing chamber temperatures. The immersion cell was empty for this set of measurements. The fitted curves (dashed lines, fit with a temperature broadened peak function) agree well with the data (colored circles) and are used to extract the electronic temperature $T_\mathrm{e}$, indicated on the side. Each peak is the result of the average over $\sim 50$ measurements. The peak amplitude $I_\mathrm{0}$ is plotted against $1/T_\mathrm{e}$ in the inset and follows the linear behavior expected for single-level transport.}
\end{figure}

Analysis of the Coulomb resonance measured with the new setup yields an electronic temperature $T_\mathrm{e}=\mathrm{\SI{9.49\pm{}0.03}{\milli\kelvin}}$. We sampled and fitted the peak several times to obtain a statistically meaningful result [Fig.~\ref{fig4}(a)]. The lower input voltage noise of the two-stage IV-converter yields a significantly reduced temperature of the electron gas as compared to the result obtained in step S2. This demonstrates the impact of the measurement electronics on state-of-the-art cryogenic setups.

Next, we applied heating power to the mixing chamber of the refrigerator and performed a temperature-dependent analysis of the conductance of the \QD{}. Figure~\ref{fig4}(b) shows the results of this experiment. The peak broadens and its amplitude $I_\mathrm{0}$ decreases with increasing temperature, as shown in the inset of Fig.~\ref{fig4}(b): $I_\mathrm{0}$ linearly depends on $1/T_\mathrm{e}$, behavior expected for a \QD{} in the single-level transport regime. Using the slope of this linear dependence and having tuned the dot such that coupling to the source and drain leads is symmetric, we extract the tunnel coupling $\Gamma =\mathrm{\SI{0.04}{\micro\electronvolt}}$. With this tunnel coupling, the full width at half maximum (FWHM) of a Lorentzian peak $2\Gamma =\mathrm{\SI{0.08}{\micro\electronvolt}}$ is much smaller than the FWHM of a thermally broadened resonance $3.55k_\mathrm{\textsc{b}}T_\mathrm{e} = \mathrm{\SI{2.90}{\micro\electronvolt}}$. This confirms that temperature broadening dominates over tunnel coupling broadening and justifies the use of Eq.~(\ref{eq:peak}).

Up to step S3 the immersion cell has been kept empty. The improvements to the electronic temperature achieved so far have been obtained by working on the room-temperature part of the electronics. It is possible to observe its influence only thanks to a well-optimized measurement setup with respect to the choice of materials, thermal anchoring, cabling, shielding and so forth. The measurements discussed for the remaining steps (S4 to S6), are performed with the immersion cell filled with superfluid $\mathrm{^4He}$.

\subsection{\label{ssec:immersion}Immersion cell and measurement parameters optimization}

We proceeded with filling $\mathrm{^4He}$ into the immersion cell. During this process, the temperature of the mixing chamber plate increases to around \mbox{\SIrange[range-phrase = -- , range-units = single]{300}{500}{\milli\kelvin}} due to the additional heat load of the gas in the capillary line and to the condensation of $\mathrm{^4He}$ in the cell. Then we empty the filling line to remove the thermal link created by the superfluid creep of $\mathrm{^4He}$. Afterwards the refrigerator cools down to base temperature, as described in section~\ref{ssec:filling}. The filling procedure did not significantly alter the quantum dot's tuning and only minimal adjustments of the applied voltages were necessary.

\begin{figure*}
\includegraphics{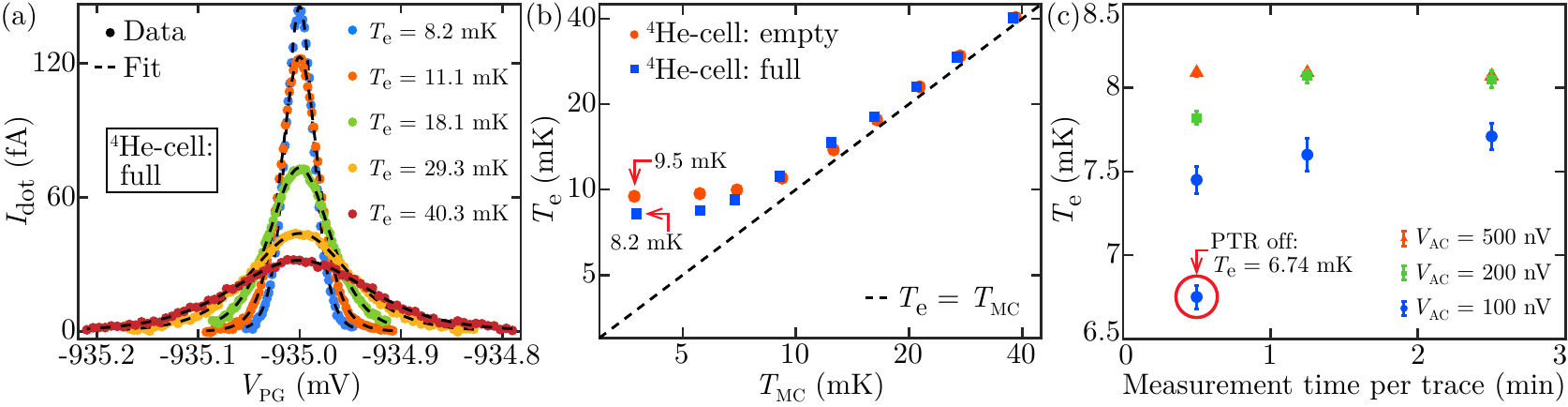}
\caption{\label{fig5}(a) Dot current $I_\mathrm{dot}$ as a function of the plunger gate voltage $V_\mathrm{\textsc{pg}}$ measured at different temperatures, similar to Fig.~\ref{fig3}(b), but with the immersion cell filled with superfluid $\mathrm{^4He}$ (step S4, see table~\ref{table}). (b) Electronic temperature $T_\mathrm{e}$ versus mixing chamber temperature $T_\mathrm{\textsc{mc}}$, both for the case of empty and full immersion cell. The dashed line represents the case of $T_\mathrm{e} = T_\mathrm{\textsc{mc}}$. In both cases, $T_\mathrm{e}$ matches the mixing chamber temperature in the limit of high $T_\mathrm{\textsc{mc}}$. On the other hand, the thermalization of the electron gas to the cooling power of the cryostat is improved in the case of immersion in superfluid $\mathrm{^4He}$. (c) Dependence of the electronic temperature $T_\mathrm{e}$ on the ac excitation $V_\mathrm{\textsc{ac}}$ applied to the system and on the duration of a single measurement of the Coulomb resonance in the dot current $I_\mathrm{dot}$ (step S5, see table~\ref{table}). We obtain each data point following the same procedure as in Fig.~\ref{fig4}(a). Decreasing both $V_\mathrm{\textsc{ac}}$ and the duration of the measurement results in a lower $T_\mathrm{e}$. The data point circled in red has been measured with the pulse tube compressor of the setup turned off (step S6).}
\end{figure*}

In step S4, we measured again the temperature dependence of the conductance of the dot [Fig.~\ref{fig5}(a)]. Measurements were performed under the same conditions and with the same electronics as in step S3. We measured a lowest electronic temperature $T_\mathrm{e}=\mathrm{\SI{8.2\pm{}0.04}{\milli\kelvin}}$, reduced with respect to the case of the empty cell. We confirmed again the single-level transport regime for the \QD{} and that the tunnel coupling $\Gamma =\mathrm{\SI{0.05}{\micro\electronvolt}}$ is negligible compared to temperature (not shown).

The effect of the superfluid $\mathrm{^4He}$ is visible when comparing $T_\mathrm{e}$ measured both with the empty and full cell against the mixing chamber temperature $T_\mathrm{\textsc{mc}}$, as shown in Fig.~\ref{fig5}(b). At higher $T_\mathrm{\textsc{mc}}$, measurements under both conditions match each other, even though they do not perfectly match the temperature measured at the mixing chamber of the refrigerator. At the lowest temperatures however, $T_\mathrm{e}$ saturates at lower values when the sample is immersed in $\mathrm{^4He}$. We understand this by considering that the thermal conductance of metals and electrically insulating materials typically decreases at least linearly with temperature.\cite{ekin_experimental_2006} The superfluid $\mathrm{^4He}$ bath mitigates this issue by providing additional thermal coupling between the elements in the immersion cell.

The mismatch between $T_\mathrm{e}$ and $T_\mathrm{\textsc{mc}}$ at higher temperatures is more puzzling. Our hypothesis is that due to some thermal impedance between the bottom part of the silver tail and the mixing chamber plate, the former remains warmer than intended. The soldering between different parts of the tail is a possible cause for this: if not enough force is applied between the connected parts, then the interfaces can give some non negligible residual thermal impedance,\cite{ekin_experimental_2006} especially at the lowest temperatures. Another possibility is that the sample does not thermalize perfectly to the tail. Radiation heating non-perfectly shielded from higher temperature stages or heating due to noise transferred via the measurement lines together with finite thermal impedance could explain the presence of a temperature gradient. Improving on this would possibly lead to a further decrease in electronic temperature but requires the production and the assembling of new parts of the low-temperature side of the setup.

Even though so far our quantum dot thermometry has benefited from improvements to the setup, this technique eventually becomes limited by the excitation applied to the sample necessary for reaching a decent signal-to-noise in the measurements and by the charge noise present in the device itself. Electrons fluctuate in and out of charge traps around the dot and in this way uncontrollably alter the electrostatic landscape felt by the dot electrons in time. As a result the position of the peak fluctuates over time and this charge noise is picked up during the time necessary to sample the Coulomb resonance.

In step S5 of our optimization procedure, we investigated the effect of both the ac excitation amplitude $V_\mathrm{\textsc{ac}}$ and the duration of the measurement on the estimated electronic temperature $T_\mathrm{e}$. Figure~\ref{fig5}(c) shows the result of this experiment: we observe that a reduction in the measured $T_\mathrm{e}$ is achieved by applying a smaller $V_\mathrm{\textsc{ac}}$. The remarkable signal-to-noise ratio achieved with our setup allows us to resolve the Coulomb peak even with an excitation as low as \SI{100}{\nano\volt}, a range where the improvements in $T_\mathrm{e}$ is even more pronounced.

Even more remarkable is the improvement achieved simply by tuning the duration of the measurements. We leave the time constant and the filter steepness of the lock-in amplifier unchanged, such that the time to acquire a single data point is fixed. The waiting time between the acquisition of two points is \SI{500}{\milli\second}. It follows that the duration of the measurements depends on how many points are acquired during the sweeping of $V_\mathrm{\textsc{pg}}$ to sample the conductance resonance over a fixed $V_\mathrm{\textsc{pg}}$ range. Initially we acquired $250$ points per trace, resulting in a measurement time of about \SI{2.5}{\minute} per trace. The values estimated for $T_\mathrm{e}$ were then averaged over $50$ measured traces. For this experiment we reduced the number of points per trace to $125$ or $50$, resulting in measurement times of \SI{1.25}{\minute} and \SI{0.5}{\minute} respectively. To compensate for the lower number of points per trace, we increased the number of traces used for the averaging procedure of $T_\mathrm{e}$ to 100 and 250 respectively, in order to keep the total number of acquired data points constant.

As seen in Fig.~\ref{fig5}(c), the estimates for $T_\mathrm{e}$ are consistently lower when the measurement time per trace is reduced. A shorter duration of the measurements results in an effective high-pass filtering on the low-frequency spectrum of the noise picked up in the measurements. Faster measurements reduce the impact of slower drifts and fluctuations due to charge noise originated in the device and to instrumentation noise, thus minimizing their effect on the sampled peak. Measurements performed with the lowest number of points that allowed us to perform a trustworthy fit to the conductance resonance and with the lowest ac excitation that gave us decent signal-to-noise ratio yield the lowest temperature $T_\mathrm{e}=\mathrm{\SI{7.45\pm{}0.08}{\milli\kelvin}}$. This is the lowest temperature measured in our setup under normal operation conditions of the refrigerator, i.e. pulse tube compressors and dilution unit running.

We observed a further reduction in the temperature $T_\mathrm{e}$ by reducing the mechanical vibrations transmitted to the refrigerator. Pulse tube compressors are necessary for the continuous operation of the cryostat as they remove heat from the high temperature stages of the setup and are responsible for the cooling of the magnet coil. Turning them off causes an almost immediate increase in the temperature of the coil and of the part of the cryostat directly cooled by the pulse tube compressors. However, the mixing chamber plate retains its temperature for a sufficient time to perform a limited number of measurements.

In step S6, we probed the electronic temperature of the device while keeping the setup in this temporary working condition. We measured the record temperature for our laboratories $T_\mathrm{e}=\mathrm{\SI{6.74\pm{}0.07}{\milli\kelvin}}$, showing how mechanical vibrations significantly influence a strongly optimized measurement setup [Fig.~\ref{fig5}(c)]. The effect of vibrations is an even greater hindrance in a magnetic field, where additional heating is caused by eddy currents. We believe unwanted induced voltages due to moving cables enclosing a time-varying magnetic flux not to be a relevant problem in our setup thanks to the structure of the flatband measurement lines.

The sequence of steps followed during this experiment was intended to first optimize as much as possible the setup without involving the $\mathrm{^4He}$ cell. During steps S1 to S3, we optimized the room temperature filtering and measurement instrumentation. We wanted to then gauge the influence of superfluid $\mathrm{^4He}$ on a setup already optimized and check how advantageous this method would be for our experiments (step S4). The additional decrease in $T_\mathrm{e}$ observed during step S5 was somehow unexpected. We observed it while testing the capabilities of our low-noise amplifiers, especially in terms of the achieved signal-to-noise ratio. Given the time investment necessary to perform these experiments, we did not repeat step S5 with an empty cell, but one could argue that even without the immersion in $\mathrm{^4He}$, the optimization of ac excitation and measurement duration would be advantageous for ultra-low temperature experiments.

\section{\label{sec:conclusions}Conclusions}

Achieving low electron temperature requires careful consideration of several aspects of a measurement setup. We reported on a dilution refrigerator that reaches $T_\mathrm{e}=\mathrm{\SI{13.0\pm{}0.1}{\milli\kelvin}}$ thanks to specially designed measurement wires, thermal anchoring and custom-made low-noise electronics.

Further optimization of the measurement electronics and the use of a superfluid $\mathrm{^4He}$ immersion cell allow us to reduce $T_\mathrm{e}$ even more. We discussed the realization and operation of the immersion cell and in particular of a valve that can be operated at few millikelvin and that is tight against superleaks.

Turning off the \PTR{} and reducing the mechanical vibrations yields the lowest measured value of $T_\mathrm{e}=\mathrm{\SI{6.74\pm{}0.07}{\milli\kelvin}}$. This result has been obtained via quantum dot thermometry in the single-level transport regime and with the host material [Al]GaAs used also for the investigation of fundamental quantum phenomena, extending the validity of this measurement beyond the thermometer itself. The designs and concepts discussed here can be transferred to other commercially available cryogenic setups.

\begin{acknowledgments}
We thank G. Cs\'a{}thy and M.\ Eich for helpful discussions. We also thank T.\ B\"ahler for the drawings of the cold valve. We acknowledge support from the ETH FIRST laboratory and financial support from the National Centre of Competence in Research "QSIT - Quantum Science and Technology".
\end{acknowledgments}

\nocite{*}
\bibliography{He4Cell_LowTemp_Paper_Bibliography}

\end{document}